\documentclass{iopart}
\usepackage{iopams}

\pdfoutput=1

\usepackage{hyperref}
\usepackage{graphicx}
\usepackage{mathptmx}
\usepackage[ngerman,american]{babel}
\usepackage{cite}
\usepackage{url}
\usepackage{textcomp}
\usepackage{wasysym}

\usepackage{algorithm}
\usepackage{algpseudocode}
\usepackage{booktabs}

\newcommand{\bigo}[1]{O(#1)}
\newcommand{\E}{\mathrm{e}}
\newcommand{\False}{\textbf{false}}
\newcommand{\True}{\textbf{true}}
\newcommand{\And}{\textbf{and}}

\newcommand{\Not}{\textbf{not}}

\newcommand{\lWhile}[2]{\State{\textbf{while} #1 \textbf{do} #2}}
\newcommand{\lIf}[2]{\State{\textbf{if} #1 \textbf{then} #2}}

\newcommand{\eqref}[1]{(\ref{#1})}
\newcommand{\text}[1]{\mathrm{#1}}

\begin{document}
\title{Stable Roommates Problem with Random Preferences}

\author{Stephan Mertens$^{1,2}$}

\address{\selectlanguage{ngerman}{$^1$Institut\ f"ur\ Theoretische\ Physik,
    Otto-von-Guericke Universit"at, PF~4120, 39016 Magdeburg,
    Germany}} 

\address{$^2$Santa Fe Institute,
1399 Hyde Park Rd,
Santa Fe, NM 87501,
USA}

\ead{mertens@ovgu.de}

\begin{abstract}
The stable roommates problem with $n$ agents has
worst case complexity $O(n^2)$ in time and space. Random
instances can be solved faster and with less memory, however.
We introduce an algorithm that has average time and space complexity
$O(n^\frac{3}{2})$ for random instances.
We use this algorithm to simulate large instances of the
stable roommates problem and to measure the probabilty $p_n$ that a
random instance of size $n$ admits a stable matching. Our data
supports the conjecture that $p_n = \Theta(n^{-1/4})$.

\noindent{\it Keywords}: analysis of algorithms, typical-case computational
  complexity, interacting agent models, disordered systems (theory)

\vspace{2ex}
\noindent
Published in: \textit{J. Stat. Mech.} (2015) P01020
\end{abstract}



\section{Introduction}

Matching under preferences is a topic of great practical importance,
deep mathematical structure, and elegant algorithmics
\cite{manlove:book,gusfield:irving:book}. The most famous example is the stable marriage problem,
where $n$ men and $n$ women compete with each other in the ``marriage
market.'' Each man ranks all the women according to his individual
preferences, and each woman does the same with all men. Everybody
wants to get married to someone at the top of his or her list, but
mutual attraction is not symmetric and frustration and compromises are
unavoidable. A minimum requirement is a matching of
men and women such that no man and woman would agree to
leave their assigned partners in order to marry each other. Such a
matching is called stable since no individual has an icentive to break
it. The problem then is to find such a stable matching.

The stable marriage problem was introduced by David Gale
and Lloyd Shapley in 1962 \cite{gale:shapley:62}. In their seminal
paper they proved that each instance of the marriage problem has at
least one stable solution, and they presented an efficient algorithm
to find it.  Since then, the Gale-Shapley algorithm has been applied to many
real-world problems, not by dating agencies but by central bodies that
organize two-sided markets like the assignment of students to
colleges or residents to hospitals \cite{roth:sotomayor:book}. 
Besides its practical relevance, the stable marriage problem has many interesting
theoretical features that have attracted
researchers from various field, including physics
\cite{omero:etal:97,nieuwenhuizen:98,dzierzawa:omero:00,
  caldarelli:capocci:01,lage:mulet:05}.

The salient feature of the stable marriage problem is its bipartite
structure: the agents form two groups (men and women), and matchings
are only allowed between these groups but not within a group.  This is
adequate for two-sided markets. But what about one-sided markets, like
the formation of cockpit crews from a pool of pilots or the assignment
of students to the double bedrooms in a dormitory? The latter is 
known as the stable roommates problem. It is the paradigmatic
example for matchings in one-sided markets.

The stable roommates problem was also introduced by Gale and
Shapley \cite{gale:shapley:62}. They noted an intriguing difference
between the marriage and the roommates problem: Whereas the former
always has a solution, the latter may have none.

The Gale-Shapley algorithm for bipartite matching does not work for
non-bipartite problems like the stable roommates problem.  In fact
some people believed that the roommates problem was NP-complete
\cite{noc}, but more than 20 years after the Gale-Shapley paper, Robert
Irving presented a polynomial time algorithm for the stable roommates
problem \cite{irving:85}. Irving's algorithm either yields a stable
solution or ``No'' if none exists.

An instance of the stable roommates problem consists of an even number
$n$ of persons (students, pilots), each of whom ranks all of the
others in strict order of preference. Since each person has to keep a
list of preferences for all $n-1$ other persons, an instance of the
stable roommates problem has size $\Theta(n^2)$. Irving's algorithm
has time complexity $O(n^2)$. This is optimal if we assume that one
has to look at the complete instance (or at least a finite fraction of
it) in order to solve the problem.

In this paper we show that in random instances, Irving's algorithm
only looks at $O(\sqrt{n})$ entries in each preference list, and we
provide a modification of the algorithm that has average
time and space complexity $O(n^{\frac{3}{2}})$. We use this
algorithm to compute the probability
$p_n$ that a random instance of size $n$ has a solution for
systems that are more than $500$ times larger than previously
simulated systems \cite{mertens:matchings}.

The paper is organized as follows. We start with a review of
Irving's algorithm. In Section \ref{sec:method} we discuss the
complexity of Irving's algorithm for random instances and our
modification that reduces the average time and space complexity from
$O(n^2)$ to $O(n^{\frac{3}{2}})$. Section~\ref{sec:application}
comprises the results of the simulations on $p_n$, obtained with the
modified algorithm.


\section{The Algorithm}
\label{sec:algorithm}

Irving's algorithm can be expressed as a sequence of ``proposals''
from one person to another. 
If person $x$ makes a proposal to person $y$ (to share a room, to form
a cockpit crew etc.), $y$ can accept or reject this proposal. If $y$
accepts the proposal, $x$ becomes
\emph{semiengaged} to $y$. If $y$ later receives another proposal from
someone he prefers to $x$, he will accept the new proposal and cancel
the semiengagement from $x$, who will in turn look for someone else to
propose to.

As the name suggests, semiengagement is not symmetric: if $x$ is
semiengaged to $y$, $y$ can be semiangaged to $z\neq x$ or to no one.
If all semiengagements are symmetric, they represent a matching. 

Irving's algorithm proceeds in two phases. Phase I sets up
semiengagements for everybody. In phase II, these semiengagements
are modified by cyclically swapping partners until all semiengagements 
are symmetric, i.e., until they represent a matching. 
The corresponding sequence of proposals (and breakups) 
is organized such that the resulting matching is stable. If the
instance admits no stable matching, this is recognized either in phase
I or phase II by running out of partners to propose to.  

For the time being, we assume that the preferences of all participants
are stored in two 2-dimensional arrays:
\begin{itemize}
\item person[$x$,$i$]: person on position $i$ in $x$'s list,
\item rank[$x$,$y$]: position of person $y$ in $x$'s list.
\end{itemize}
The two arrays are not independent, of course, but the redundancy
allows us to look up persons and ranks in time $O(1)$. 

For random instances we initialize the preference list of person $x$ a by random
permutation of all other persons (including $x$) and then move $x$ to
the very end of its own preference list. This means that we allow 
$x$ being matched with himself as the worst choice.
If this really happens, this means that $x$ has no proper partner, i.e.
that no stable matching exists for that instance.

In our implementation of the algorithm we will access the preference lists only through
the function
\begin{algorithmic}[]
  \Function{GetData}{$x$, $i$}
     \State{$y$ := person[$x$, $i$]}
     \State{$r$ := rank[$y$, $x$]}
     \State{\Return{($y$, $r$)}}
  \EndFunction
\end{algorithmic}
which returns the pair $(y,r)$ where $y$ is the person with rank $i$ in $x$'s prefence list
and $r$ is rank of person $x$ in $y$'s preference list.

We will describe both phases of Irving's algorithm
without proving their correctness. For the proofs we refer the reader to
Irving's original paper \cite{irving:85}.

\subsection{Phase I}

Phase I of the algorithm tries to establish semiengagements for every
person. The general idea is that the first proposal of $x$ goes to the first
person on his preference list, and only if this proposal is rejected
(immediately or subsequently), $x$ proposes to the second person on
his preference list and so on. On the receiving side, $y$ accepts a
proposal only if the proposing person ranks higher on his preference
list than the person whose proposal he has currently accepted.  

Imagine the list of preferences written horizontally left (most desired partner) to right
(least desired partner). Then the proposals \emph{made} move from left to
right, while the proposals \emph{accepted} move from right to left. In
a matching, both types of proposals meet at the same position.
This motivates the names for the following lists that hold the
current set of proposals:
\begin{itemize}
\item leftperson[$x$]: the person whom $x$ is currently proposing to,
\item leftrank[$x$]: the rank of that person in $x$'s preference list,
\item rightperson[$x$]: the person from which $x$ is currently holding
  a proposal
\item rightrank[$x$]: the rank of that person in $x$'s preference list.
\end{itemize}
Again these lists are not independent, but the redundancy allows a
faster lookup especially in phase II.

With this lists, the semiengagement of $x$ to $y$ is expressed by
the simultaneous validity of the identities
$y=\mbox{leftperson}[x]$ and $x=\mbox{rightperson}[y]$. 

\begin{algorithm}
\begin{algorithmic}[1]
  \Procedure{Phase\_I}{}
  \For{$x := 1,\ldots,n$}\Comment{initialization}
      \State{$\text{holds\_proposal}[x] := \False$}\Comment{no one is
        holding a proper proposal...}
      \State{$\text{rightperson}[x] := x$}\Comment{...but a proposal from self}
      \State{$\text{rightrank}[x] := n$}\Comment{self proposals are
        the worst}
       \State{$\text{leftrank}[x] := 1$}
  \EndFor
  \For{$x := 1\ldots n$}
    \State{proposer := $x$}
    \Repeat
      \State (next, rank) := \Call{GetData}{proposer, leftrank[proposer]}
      \While{rank $>$ rightrank[next]} 
         \State leftrank[proposer] := leftrank[proposer]+1 
         \State (next, rank) := \Call{GetData}{proposer, leftrank[proposer]}
      \EndWhile
      \State{previous := rightperson[next]}
      \State{rightrank[next] := rank}
      \State{rightperson[next] := proposer}
      \State{leftperson[proposer] = next}
      \State{proposer := previous}
    \Until {holds\_proposal[next] = \textbf{false}}
    \State holds\_proposal[next] := \textbf{true}
    \If {leftrank[proposer] = $n$} \Comment{proposer engaged to himself...}
       \State{\Return \False} \Comment{... means: no stable matching possible}
    \EndIf
  \EndFor
  \State{\Return \True}
  \EndProcedure
\end{algorithmic}
\caption{Phase I of the stable roommates algorithm}
\label{alg:phase-I}
\end{algorithm}

Algorithm~\ref{alg:phase-I} shows the pseudocode for phase I of
Irving's algorithm. It stops, when every person \emph{holds} a proposal,
which implies that every person has also \emph{made} a proposal that has been
accepted, i.e. that every person is semiengaged. It returns \textbf{false} if someone has run out of partners
(and is therefore engaged to himself), which means that there is no
stable matching for this instance. If it returns \textbf{true}, we
can still hope to find a stable matching in phase II of the algorithm. 

\subsection{Phase II}

Phase I usually ends with everybody semiengaged to someone, but with
asymmetric engagements $\text{leftrank}[x] < \text{rightrank}[x]$ for most 
persons $x$. Such persons have to
give up their current proposal to leftperson[$x$] and find somebody down the list that
would (temporarily) accept a proposal from $x$. We keep track of these second choices in
the following lists
\begin{itemize}
\item secondperson[$x$]: $x$'s next best person that would
  accept a proposal
\item secondrank[$x$]: the rank of that person in $x$'s preference list,
\item secondrightrank[$x$]: the rank of $x$ in the preference list of
  secondperson[$x$].
\end{itemize}
If $x$ withdraws his proposal and proposes to $y =
\text{secondperson}[x]$, who temporarily accepts the proposal, the
previous partners of $x$ and $y$ both loose their semiengagements and
have to look themselves for their second best partners and so on. This
avalanche of break-ups and new propsals is called a rotation. It reduces the difference
rightrank[$x$]-leftrank[$x$] for several persons $x$ and is a step
towards a matching.

The key idea of phase II is to organize this rearrangement of
semiengagements in a so called all-or-nothing cycle. This is a sequence
$a_1\ldots,a_r$ of persons such that $a_i$'s current second choice is
$a_{i+1}$'s current first choice for $i=1\ldots,r-1$, and $a_r$'s
current second choice is $a_1$'s current first choice. In terms of our
lists, an all-or-nothing cycle is given by
\begin{eqnarray*}
  \text{secondperson}[a_i] &=& \text{leftperson}[a_{i+1}] \quad
i=1,\ldots,r-1\\
  \text{secondperson}[a_r] &=& \text{leftperson}[a_{1}]\,.
\end{eqnarray*}
In phase II of Irving's algorithm, an all-or-nothing cycle is
identified and the corresponding rotation is executed.  This process
is iterated until there are no more all-or-nothing cycles (in which
case we've found a stable matching) or until someone runs out of
partners after a rotation (in which case this instance has no stable
matching).

\begin{algorithm}
  \begin{algorithmic}[1]
    \Procedure{SeekCycle}{\ }
      \For{$x := 1,\ldots,n$}
       \lIf{$\text{leftrank}[x] < \text{rightrank}[x]$}{\textbf{break}} \Comment{find
            unmatched person}
       \EndFor
      \If{$\text{leftrank}[x] \geq \text{rightrank}[x]$} \Comment{no unmatched person found}
        \State{\Return ($0,0,\emptyset$)}\Comment{return empty cycle}
      \Else \Comment{unmatched person found}
         \State{$\text{last} := 1$}
         \Repeat
            \State{$\text{cycle}[\text{last}] := x$}
            \State{last := last + 1}
            \State{$p := \text{leftrank}[x]$}
            \Repeat \Comment{find second choice of $x$}
              \State{$p := p+1$}
              \State{$(y, r) :=$ \Call{GetData}{$x$, $p$}}
            \Until{$r \leq \text{rightrank}[y]$}\Comment{$y$ would
              accept a proposal}
            \State{$\text{secondrank}[x] := p$}\Comment{store second
              choice}
            \State{$\text{secondperson}[x] := y$}
            \State{$\text{secondrightrank}[x] := r$}
            \State{$x :=
              \text{rightperson}[\text{secondperson}[x]]$}\Comment{next
              element in cycle}
         \Until{$x \in$ cycle} \Comment{cycle closed}
         \State{$\text{last} := \text{last}-1$}
         \State{$\text{first} := \text{last}-1$} \Comment{rewind to start of cycle}
         \lWhile{cycle[first] $\neq x$}{$\text{first} := \text{first}-1$} 
         \State{\Return(first, last, cycle)}
      \EndIf
    \EndProcedure
  \end{algorithmic}
  \caption{Finding an all-or-nothing cycle for phase II of the stable roommates algorithm}
\label{alg:cycle}
\end{algorithm}

Algorithm~\ref{alg:cycle} shows the pseudocode for a function that
finds and returns an all-or-nothing cycle or an empty cycle. To
compute a cycle we need to identify the person whose current first choice is
$y=\text{secondperson}[x]$.  But this person is simply given by
$\text{rightperson}[y]$, i.e., it can be found in time $O(1)$ (see
line 20 of Alg.~\ref{alg:cycle}).

\begin{algorithm}
\begin{algorithmic}[1]
  \Procedure{Phase\_II}{}
  \State{solution\_possible := \True}
  \State{solution\_found := \False}
  \While{solution\_possible \And\ \Not\ solution\_found}
    \State{(first, last, cycle) := \textsc{SeekCycle}( )}
    \If{cycle is empty}{\ solution\_found := \True}
    \Else
       \For{$x := \text{cycle}[\text{first}],\ldots,\text{cycle}[\text{last}]$}
           \State{$\text{leftrank}[x] := \text{secondrank}[x]$}
          \State{$\text{leftperson}[x] := \text{secondperson}[x]$}
          \State{$\text{rightrank}[\text{leftperson}[x]] := \text{secondrightrank}[x]$}
          \State{$\text{rightperson}[\text{leftperson}[x]] = x$}
         \EndFor
         \For{$x := \text{cycle}[\text{first}],\ldots,\text{cycle}[\text{last}]$} 
             \lIf{$\text{leftrank}[x] >
               \text{rightrank}[x]$}{sol\_possible := \False}
          \EndFor
    \EndIf
  \EndWhile
  \State{\Return solution\_found}
  \EndProcedure
\end{algorithmic}
\caption{Phase II of the stable roommates algorithm}
\label{alg:phase-II}
\end{algorithm}

Algorithm~\ref{alg:phase-II} shows pseudocode for phase II which finds
an all-or-nothing cycle, executes the corresponding rotation and
iterates this until there are no more all-or-nothing cycles or a
rotation has left a person without any partners to propose to.

The complete algorithm consists of an initialization phase (not
shown), which generates a random preference list for each person,
followed by calls to \textsc{Phase\_I} and \textsc{Phase\_II}.

\section{Analysis and Modification}
\label{sec:method}

\begin{figure}
  \centering
  \includegraphics[width=0.9\columnwidth]{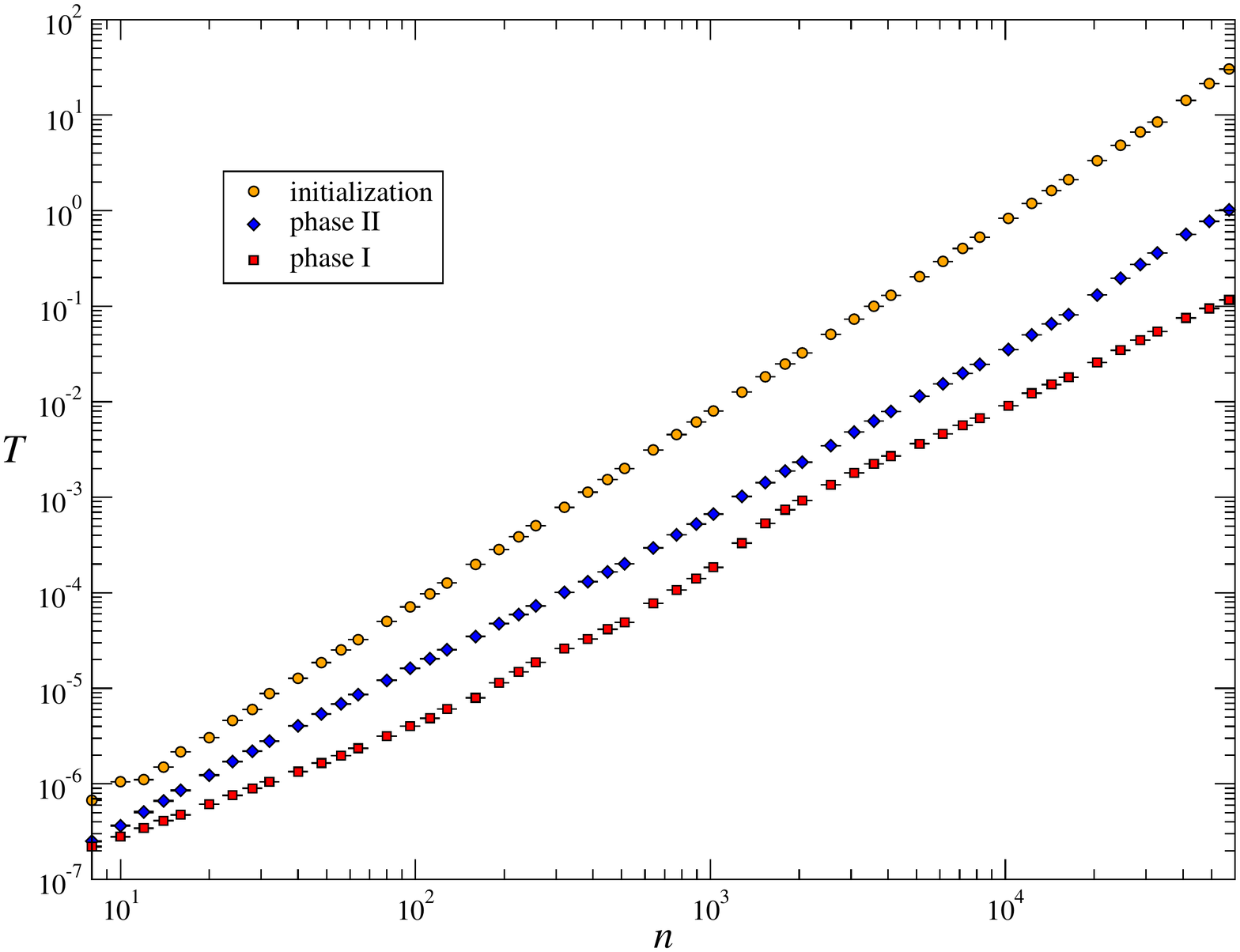}
  \caption{Running times of the various phases of Irving's algorithm
    for random instances of the stable roommates problem. The total
    running time is dominated by the time for initialization of the complete
    instance. This time scales like $n^2$, whereas the time for the
    actual solution (phases I and II) grows much slower. The time $T$
  is the average wallclock time in seconds on a single core of an Intel$^{\small\textregistered}$ Xeon$^{\small\textregistered}$ E5-1620
CPU running at 3.6 GHz. The ``bump'' in the data for phase I is
probably due to cache misses for larger systems.}
  \label{fig:benchmark}
\end{figure}

Figure~\ref{fig:benchmark} shows the average running times of the
different phases on random instances of varying size $n$. The only
phase that scales like $\Theta(n^2)$ is the initialization, i.e., the
generation of the random permutation of the preference lists. The time for the actual
solution (phase I and phase II) grows significantly slower than $n^2$,
which implies that the algorithm doesn't need to look at the complete
preference table to solve the problem.

\begin{figure}
  \centering
  \includegraphics[width=0.9\columnwidth]{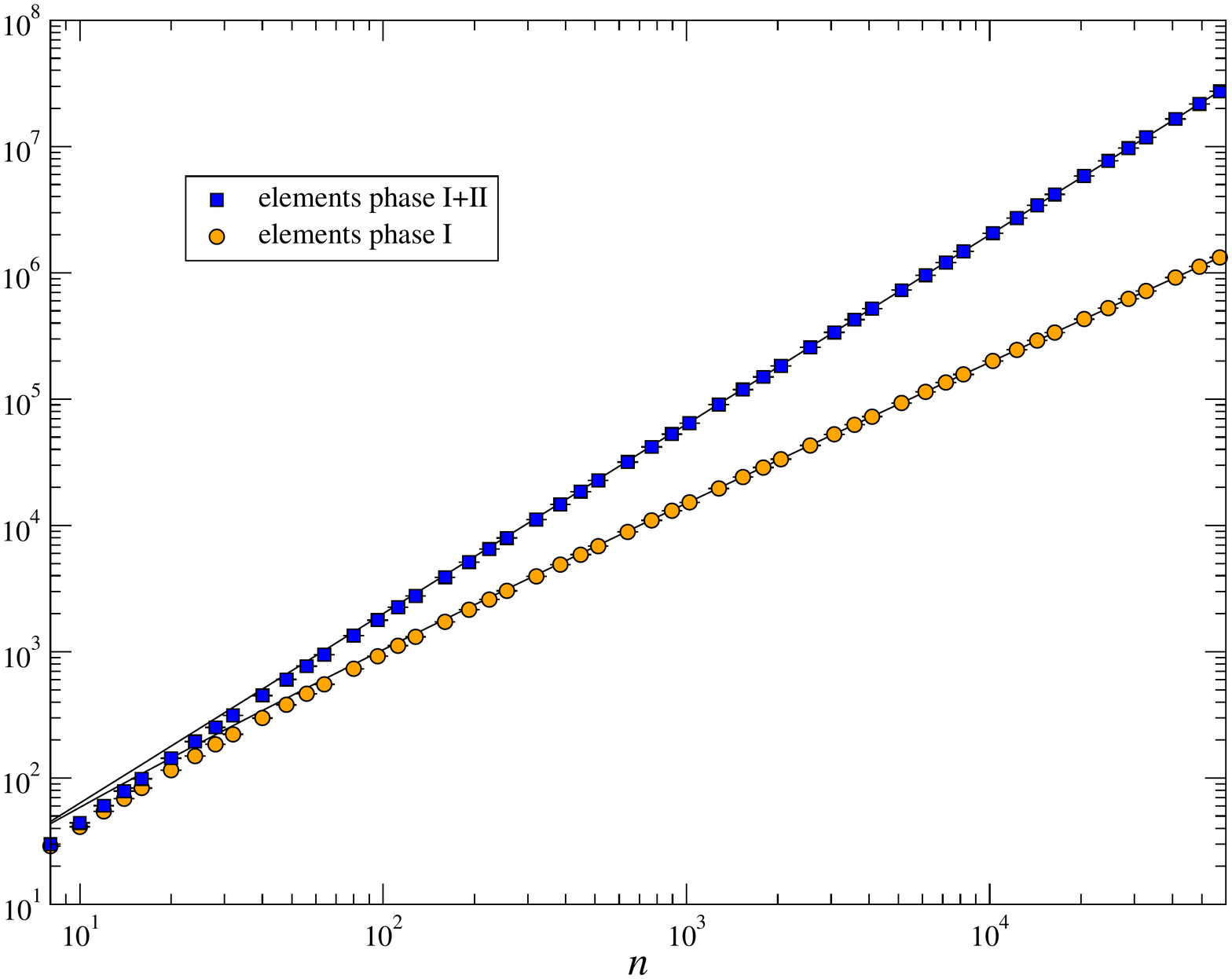}
  \caption{Number of elements in the preference tables that are
    actually read by Irving's algorithm. Each symbol
    represents an average over $10^4$ random instances. The lines are
    $2n^{\frac{3}{2}}$ for the total number of elements and $2 n H_n$ for 
    phase I, where $H_n$ is the $n$th harmonic number.   }
  \label{fig:rmi-size}
\end{figure}

Figure~\ref{fig:rmi-size} shows the average number of entries in the
preference lists that are actually read by Irving's algorithm in order
to find a stable matching or to report that no stable matching exists.
For large values of $n$, this number is $2 n^{\frac{3}{2}}$. This can
be understood by the following simple, albeit non rigorous
consideration.  Let $k$ be the average number of proposals that a
person makes in the course of Irving's algorithm. Then $k$ is also the
average number of proposals that a person receives, and the total
number of entries in the preference table involved is $2kn$.  Now
$\mbox{leftrank}[x]$ increases by one with each proposal made by $x$,
hence $\mbox{leftrank[$x$]} = k$ on average. The value of
rightrank[$x$] is given by the \emph{minimum} of $k$ values drawn
uniformly from $1,\ldots,n$. Hence the distribution of
$\mbox{rightrank}[x]=\ell$ is
\begin{displaymath}
  P(\ell) = \frac{{{n-\ell}\choose{k-1}}}{{n\choose k}}\,,
\end{displaymath}
with mean value $\frac{n+1}{k+1}$.  The algorithm terminates if
$\mbox{leftrank}[x] \simeq \mbox{rightrank}[x]$ or $k \simeq
\frac{n+1}{k+1}$. Hence $k\simeq\sqrt{n}$, and the total number of
entries read by Irving's algorithm is $2 n^{\frac{3}{2}}$. Note that
this consideration ignores the fluctuations in rightrank[$x$]: if
$k\simeq \sqrt{n}$, the standard deviation of $P(\ell)$ is
$O(\sqrt{n})$, hence the number of proposals received by an individual
person $x$ can differ considerably from the mean value $\sqrt{n}$. We
do have the strict equality between the total number of proposals made
and total number of proposals received, however. But a person, who has
\emph{received more} proposals than average, at termination has
\emph{made less} proposals than the average (and vice versa). Hence
the individual fluctuations cancel in the total number of proposals if
we assume, that at termination
$\mbox{leftrank}[x]\simeq\mbox{rightrank}[x]$ for almost all persons
$x$. This is not obvious, since in most cases (see next section) the
algorithm terminates when the first person runs out of partners to
propose to, i.e. the first time that $\mbox{leftrank}[x] >
\mbox{rightrank}[x]$ for some $x$. It could well be that at that moment the gap between
leftrank and rightrank is still large for some or many other
persons. Hence the crucial assumption that underlies our argument is
the assumption of a certain uniformity of the decrease of
$\mbox{rightrank}[x]-\mbox{leftrank}[x]$ over all persons $x$.
The fact that the observed total number of proposal is in fact narrowly
concentrated around the predicted value (Figure ~\ref{fig:rmi-size}) is an
indication that this assumption is justified.

The number of elements read in phase I is even smaller. Phase I
terminates if every person holds a proposal. Consider the sequence
$x_1,x_2,\ldots$ of persons that receive a proposal. Phase I
terminates if this sequence contains every person at least once. If we
assume that the $x_i$'s are independent random variables, uniformly
drawn from $\{1,\ldots,n\}$, this problem is known as the coupon
collector's problem \cite{noc}: an urn contains $n$ different coupons,
and a collector draws coupons from that urn with replacement. How many
coupons does the collector need to draw (on average), before he has
drawn each coupon at least once? It is well known that the collector
should expect to draw
\begin{displaymath}
  n\,\left(1 + \frac{1}{2} + \frac{1}{3} + \cdots + \frac{1}{n}\right)
  = n H_n
\end{displaymath}
coupons in order to own at least one coupon of every kind. $H_n$ is
known as the $n$th harmonic number. Note that  
\begin{equation}
  \label{eq:harmonic}
  n H_n = n\,\log n + \gamma n + \frac{1}{2} + \bigo{n^{-1}}\,,
\end{equation}
where $\gamma = 0.5772156\ldots$ is the Euler-Mascheroni constant.

In our case, the $n$ coupons are the proposals to the $n$ different
recipients, and phase I is the coupon collector. Hence the expected
number of proposals in phase I is $n H_n$, and since each proposal
implies two accesses to the preference lists, the number of elements
read in phase I should be $2n H_n$.  This is in fact the observed
asymptotic scaling, as can be seen in Figure~\ref{fig:rmi-size}.

\begin{algorithm}
\begin{algorithmic}[1]
  \Function{GetData}{$x$, $i$}
     \If{$i \in$ rank[$x$]}\Comment{element is known}
       \State{$y$ := person[$x$]($i$)}
       \State{$r$ := rank[$y$]($x$)}
     \Else  \Comment{element is new}
        \Repeat 
           \State{$y$ := random}\Comment{generate random person}
        \Until{$y \not\in$ person[$x$]}
        \Repeat 
           \State{$r$ := random}\Comment{generate random rank}
        \Until{$r \not\in$ rank[$y$]}
        \State{$\text{person}[x] := \text{person}[x]\cup
          (y,i)$}\Comment{save new table entries}
        \State{$\text{rank}[x] := \text{rank}[x]\cup (i,y)$}
        \State{$\text{person}[y] := \text{person}[y]\cup
          (x,r)$}
        \State{$\text{rank}[y] := \text{rank}[y]\cup (r,x)$}
     \EndIf
     \State{\Return{($y$,$r$)}}
  \EndFunction
\end{algorithmic}
\caption{This version of the GetData function generates the entries of
  the preference lists ``on the fly'' as they are requested. The
  preference list of person $x$ is maintained in the associative
  arrays person[$x$] and rank[$x$].}
\label{alg:GetData}
\end{algorithm}

We can exploit the fact that Irving's algorithm looks only at
$\bigo{n^{3/2}}$ elements of preference table by generating and
storing only the elements that are requested by the algorithm. This
saves us the expensive initialization phase and reduces the memory
consumption considerably. Algorithm~\ref{alg:GetData} shows the
corresponding version of the function \textsc{GetData}. It maintains
two arrays (person and rank) of maps. A map (aka associative array) is
a data structure that holds pairs $(k, v)$ where $k$ is the key and
$v$ is the value of the data element.  In the map person[$x$], the key
is the rank and the value is the person of that rank in $x$'s
preference list. The map rank[$x$] holds the same data elements but
with the role of key and value reversed.  The rationale behind this
redundancy is again efficiency: using hash tables, a map can be
implemented such that the lookup of a value given the key can be done
in (expected) constant time, independent of the number of elements.
Hence \textsc{GetData} has average time complexity $\bigo{1}$ when the
requested data element is already known. When the requested element is
new, the generation of a new random element may take longer since we
need to generate a random person $y$ that is not contained in $x$'s
preference list so far and an unoccupied rank $r$ for person $x$ in
$y$'s list.  Both are computed by a simple loop that generates random
numbers until it hits a number not already contained in the list. In
our case this is a reasonable approach since we know that the expected
number of elements in person[$x$] and rank[$y$] is $\bigo{\sqrt{n}}$,
hence the expected number of iterations in our loop is
$1+\bigo{n^{-1/2}}$.  Since inserting new elements in a map can also
be done in constant time, the average time complexity of
\textsc{GetData} is $\bigo{1}$. Each map is initialized with the
single entry person[$x$][$n$]=$x$ and rank[$x$][$x$]=$n$, all other
entries are only added as needed. 

\begin{figure}
  \centering
  \includegraphics[width=0.9\columnwidth]{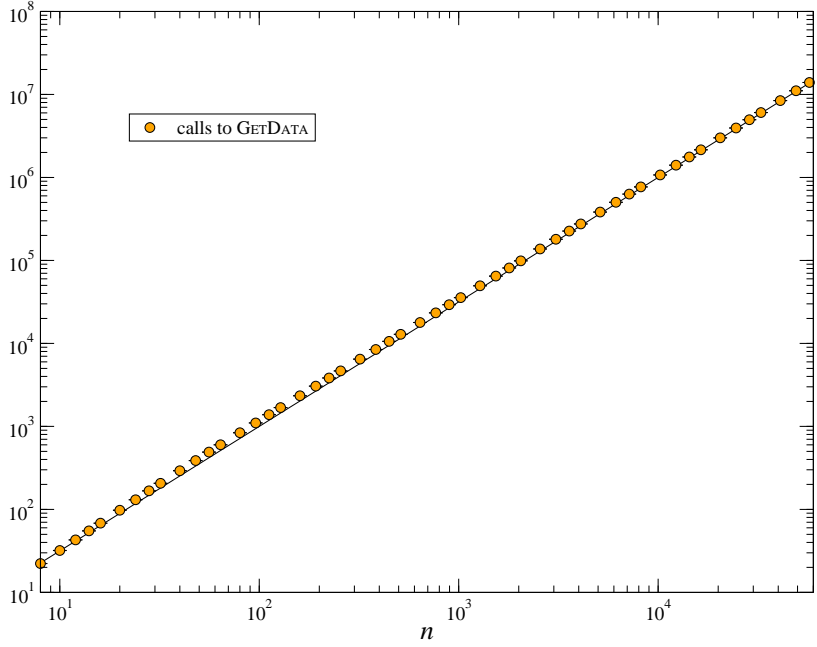}
  \caption{Number of calls of function \textsc{GetData}. Each symbol
    represents an average over $10^4$ random instances. The line is
    $n^{\frac{3}{2}}$. }
  \label{fig:rmi-calls}
\end{figure}

The function \textsc{GetData} is called unconditionally from within
the innermost loop in phase I. In phase II it is called for each
element in search for a cycle (including all cycle elements). Hence
the number of calls of \textsc{GetData} is a good measure for the
average time complexity of the algorithm. As can be seen in
Figure~\ref{fig:rmi-calls}, the average time complexity is indeed
$\Theta(n^{\frac{3}{2}})$.

\begin{figure}
  \centering
  \includegraphics[width=0.9\columnwidth]{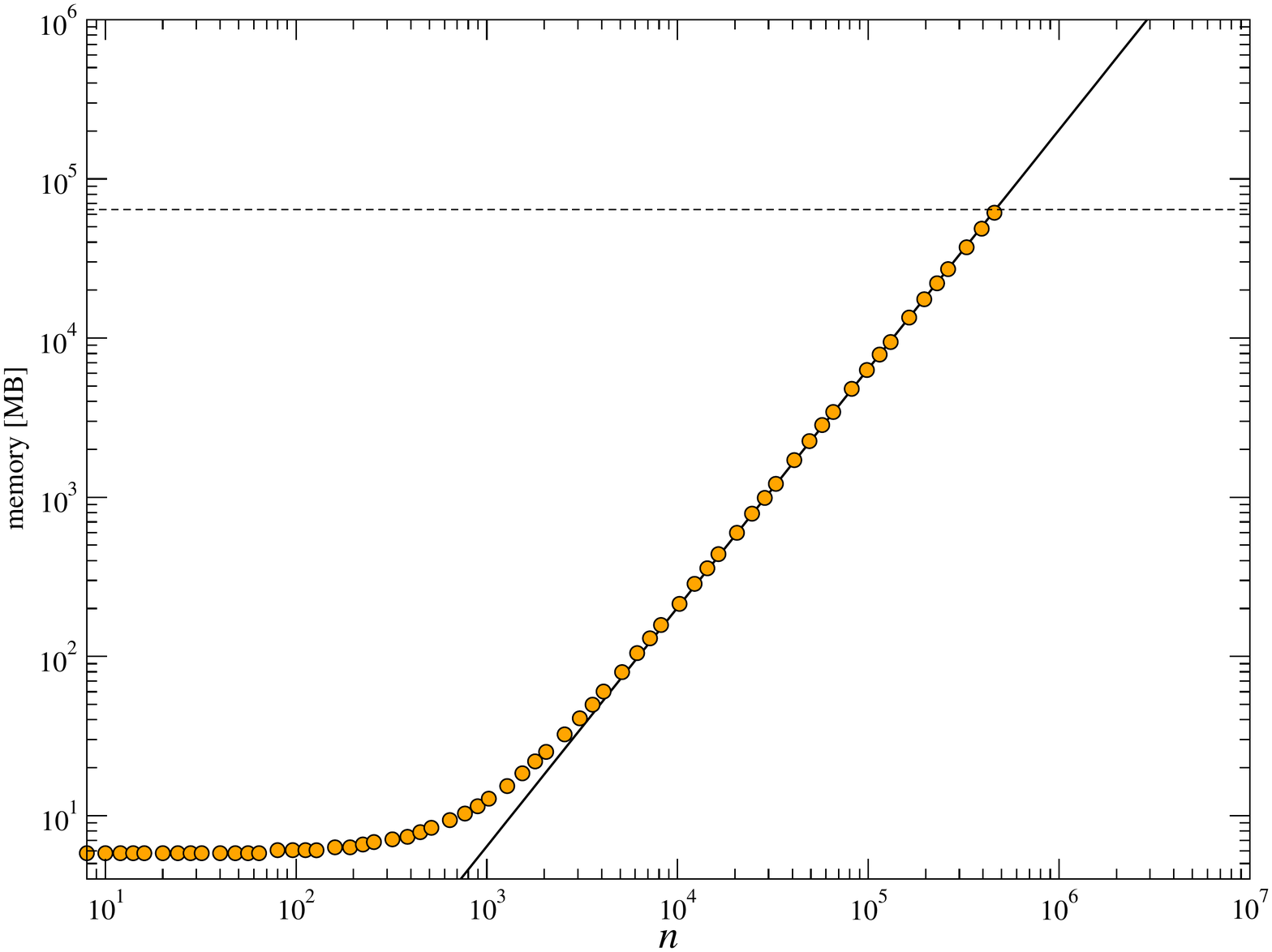}
  \caption{Actual memory consumption of Irving's algorithm for random instances, implemented with associative maps. The solid line is a numerical $\Theta(n^{3/2})$ fit, the dashed line marks the 64 GBytes limit of our hardware.  }
  \label{fig:memusage}
\end{figure}

The actual memory usage depends on the implementation of the map. We
used the container class \texttt{unordered\_map} from the C++ standard
library provided by the GNU Compiler Collection
(\url{http://gcc.gnu.org}). Figure~\ref{fig:memusage} shows the total
memory usage of the program. For large values of $n$ it is dominated
by the maps \texttt{person} and \texttt{rank} and scales like
$n^{3/2}$. On a computer with 64 GBytes of memory our implementation
allows us to run problems up to $n \simeq 470\,000$, which is more
than $5$ times larger than the maximum problem size allowed by a
straightforward implementation with storage of the complete preference
tables. The memory efficiency of our implementation is in fact so good
that time becomes the bottleneck, as we will see in the next section.

\section{Application}
\label{sec:application}

A long standing open problem is the computation of 
the probability $p_n$ that a random instance of the stable roommates
problem of size $n$ has a stable matching \cite[problem
8]{gusfield:irving:book}. In particular one is interested in the
asymptotic behavior of $p_n$ as $n$ grows large.

There is an integral representation for $p_n$ that can be used
to compute $p_n$ exactly \cite{pittel:93}. Unfortunately, 
the number of terms in the integral increases exponentially with $n^2$,
which had limited the explicit evaluation of $p_n$ to the case $n=4$. Using a computer
algebra system, we evaluated the integrals for $n \leq 10$ \cite{mertens:rmp-integral}:
\begin{eqnarray*}
   \label{eq:exactvalues}
   p_4 &=& \frac {26}{27} = 0.96296\ldots \\[2ex]
   p_6 &=& \frac {181431847}{194400000} = 0.93329\ldots\\[2ex]
   p_8 &=&  {\frac {809419574956627}{889426440000000}} =
   0.910046\ldots\\[2ex]
   p_{10} &=&
   \frac{25365465754520943457921774207}{28460490127321448448000000000}
   = 0.891251\ldots\,.
\end{eqnarray*}

Computing $p_n$ for larger values of $n$ requires Monte Carlo simulations.  
These simulations indicate that $p_n$ is a monotonically decreasing
function of $n$, but early simulations up to $n=2000$ \cite{pittel:irving:94} did not
settle the question as to whether $p_n$ converges to $0$ or to some
positive constant. The problem with simulations is that the decay of $p_n$ is rather
slow. In fact Pittel \cite{pittel:93} proved the asymptotic lower bound
\begin{equation}
  \label{eq:lower-bound}
  p_n \gtrsim \frac{2\E^{3/2}}{\sqrt{\pi n}}
\end{equation}
by applying the second moment method to the number of stable matchings.
Extended simulations \cite{mertens:matchings} up to $n=20000$ 
suggested an algebraic decay $p_n \simeq a n^{-\delta}$. 
The numerical data from \cite{mertens:matchings} was used to
boldly conjecture the values of $a$ and $\delta$ as 
\begin{equation}
  \label{eq:conjecture}
  p_n \simeq \mathrm{e}\sqrt{\frac{2}{\pi}}\,\,n^{-1/4}\,.
\end{equation}
Using our algorithm with reduced running time and memory
consumption, we can check this conjecture against
extended numerical data.

We simulated systems of size $n = n_0 2^k$,
$k=0,\ldots,k_{\text{max}}$ and $n_{0} \in\{8,10,12,14\}$ where
$k_{\text{max}}$ is limited by the available memory. In our case this
means $k_{\text{max}} \leq 15$ (Figure~\ref{fig:memusage}). The
corresponding instances have an effective size that is
more than $500$ times larger than the
largest systems investigated in \cite{mertens:matchings}.

To measure $p_n$, we generate and solve $M$ independent random
instances of size $n$ and record the fraction $\hat{p}_n$ of samples that admit a
stable matching. The 95\% confidence interval for $p_n$ is then
$\hat{p}_n \pm 2\sigma_n$, where the standard deviation $\sigma$ is given by
\begin{equation}
  \label{eq:variance}
  \sigma_n = \sqrt{\frac{\hat{p}_n(1-\hat{p}_n)}{M}}\,.
\end{equation}
We vary the number of samples $M$ with the system size $n$. 
We used values from $M=10^{10}$ for small values of $n$ down to $M=10^4$ for 
the largest values of $n$. Table~\ref{tab:results} shows the results.

\begin{table}
  \begin{center}
  \footnotesize
  \begin{tabular}{rl@{\hspace{12mm}}rl@{\hspace{12mm}}rl@{\hspace{12mm}}rl}\toprule
    $n$ & $p_n$ & $n$ & $p_n$ & $n$ & $p_n$ & $n$ & $p_n$ \\\midrule
    8 & 0.910048(5)   &    128 & 0.60986(1)   &   2048 & 0.32473(9)   &  32768 & 0.1650(4)  \\
   10 & 0.891247(6)   &    160 & 0.58183(1)   &   2560 & 0.30794(9)   &  40960 & 0.1563(4)  \\
   12 & 0.875525(6)   &    192 & 0.55946(1)   &   3072 & 0.29464(9)   &  49152 & 0.1494(3)  \\
   14 & 0.861952(6)   &    224 & 0.54099(1)   &   3584 & 0.28391(9)   &  57344 & 0.1440(5)  \\
   16 & 0.849958(7)   &    256 & 0.52536(1)   &   4096 & 0.27486(9)   &  65536 & 0.140(2)  \\
   20 & 0.829239(7)   &    320 & 0.49993(1)   &   5120 & 0.26042(9)   &  81920 & 0.132(2)  \\
   24 & 0.811499(7)   &    384 & 0.47987(1)   &   6144 & 0.2492(2)   &  98304 & 0.126(2)  \\
   28 & 0.795768(7)   &    448 & 0.46339(2)   &   7168 & 0.2402(2)   &  114688 & 0.120(2)  \\
   32 & 0.781542(8)   &    512 & 0.44949(2)   &   8192 & 0.2320(2)   &  131072 & 0.118(2)  \\
   40 & 0.756482(8)   &    640 & 0.42704(2)   &  10240 & 0.2200(2)   &
   163840 & 0.111
(2)  \\
   48 & 0.734851(8)   &    768 & 0.40939(7)   &  12288 & 0.2103(2)   &  196608 & 0.103(6)  \\
   56 & 0.715866(8)   &    896 & 0.39482(7)   &  14336 & 0.2024(3)   &  229376 & 0.104(6)  \\
   64 & 0.699044(9)   &   1024 & 0.38270(9)   &  16384 & 0.1961(3)   &  262144 & 0.098(6)  \\
   80 & 0.670377(9)   &   1280 & 0.36327(9)   &  20480 & 0.1854(4)   &
   327680 &  0.097(6)\\ 
   96 & 0.646797(9)   &   1536 & 0.34782(9)   &  24576 & 0.1774(4)   &  393216 &  0.089(6)\\
  112 & 0.62692(1)   &   1792 & 0.33526(9)   &  28672 & 0.1709(4)   & 458752 & 0.085(5)\\\bottomrule
  \end{tabular}
  \end{center}
\caption{Values of $p_n$ from simulations. Notation: $p_8 =
  0.910048(5)$ means that $[0.910048-0.000005,0.910048+0.000005]$ is the 95\%
  confidence interval for $p_8$.}
\label{tab:results}
\end{table}

We ran our simulation on a small cluster consisting of five
nodes. Each node has 64 GByte of RAM and two Intel Xeon CPU E5-2630
running at 2.30GHz. Each CPU has 6 real cores or (using
hpyerthreading) 12 virtual cores. For smaller systems, we can use all
$5\times 2\times 12 =120$ virtual cores to solve instances in
parallel, but for the larger sizes, the available memory per core
limits the number of usable cores.  The available memory allows us to compute
problems up to $n=2^{15}\times 14 = 458752$ using only one core per
node.  Solving a single instance of this size takes about 14 minutes,
i.e. solving a sample of $M=10^4$ instances as in
Table~\ref{tab:results}  takes about 20
days on $5$ nodes in parallel.

\begin{figure}
  \centering
  \includegraphics[width=0.9\columnwidth]{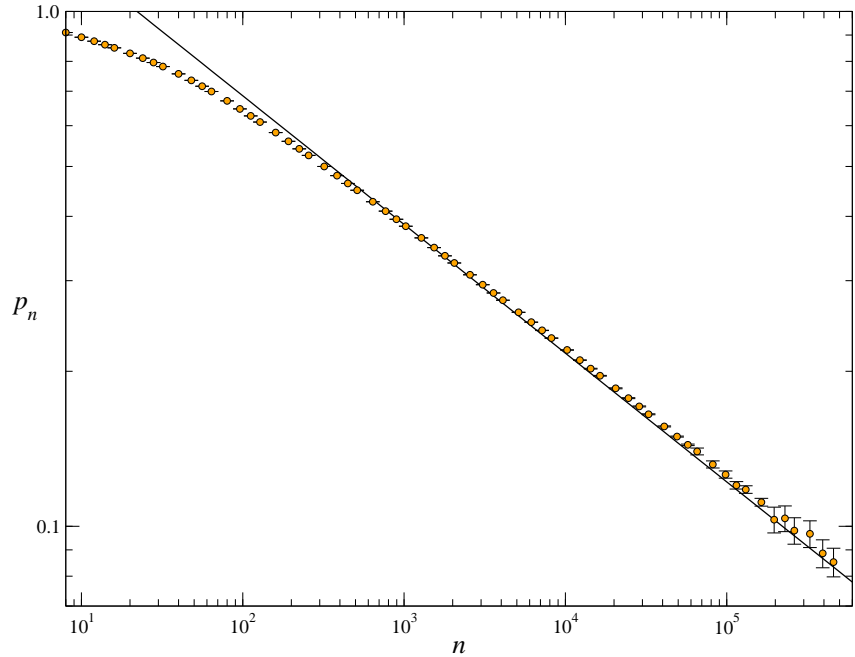}
  \caption{Simulation results for $p_n$ in a log-log-plot. Error bars denote 95\%
    confidence intervals. The line represents the conjectured
    asymptotic behavior \eqref{eq:conjecture}.}
  \label{fig:rmates}
\end{figure}

Figure~\ref{fig:rmates}
shows $p_n$ versus $n$ in a log-log-plot. The data supports an
asymptotic algebraic decay $p_n \simeq a n^{-\delta}$ for some constants
$\delta$ and $a$, in agreement with the conjecture
\eqref{eq:conjecture}, which is also displayed in
Figure~\ref{fig:rmates}. The visual impression suggests that
$\delta=1/4$ as claimed in \eqref{eq:conjecture}, but that 
the true prefactor $a$ is slightly larger than $\E\sqrt{2/\pi}$.
In fact, a least squares fit of the one-parameter function $a\,n^{-1/4}$ 
to the data points for $n\geq 32765$ yields
$a=2.223(3)$, which is 3\%\ larger than $\E\sqrt{2/\pi} = 2.1688\ldots$.

\begin{figure}
  \centering
  \includegraphics[width=0.48\columnwidth]{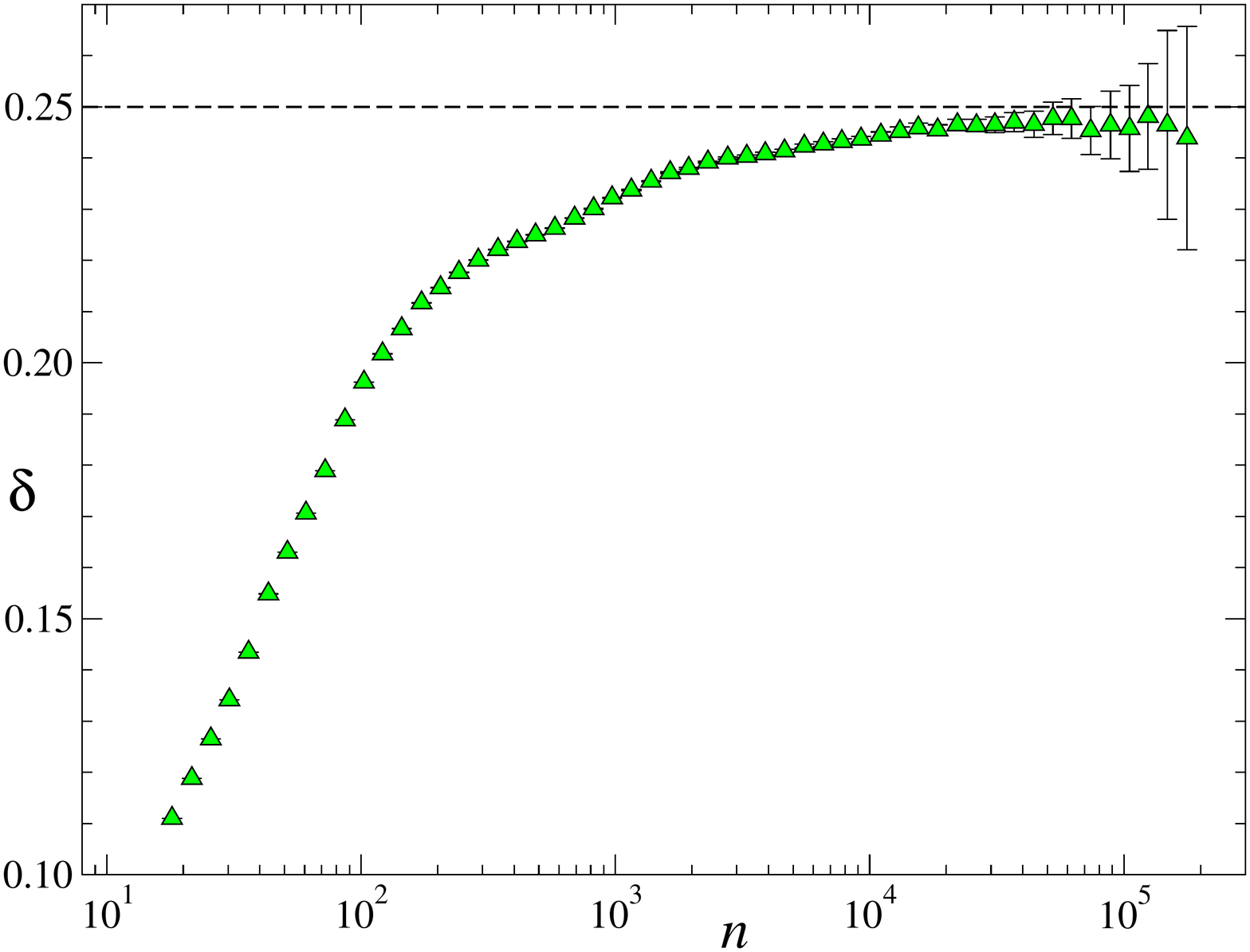} \hfill
  \includegraphics[width=0.48\columnwidth]{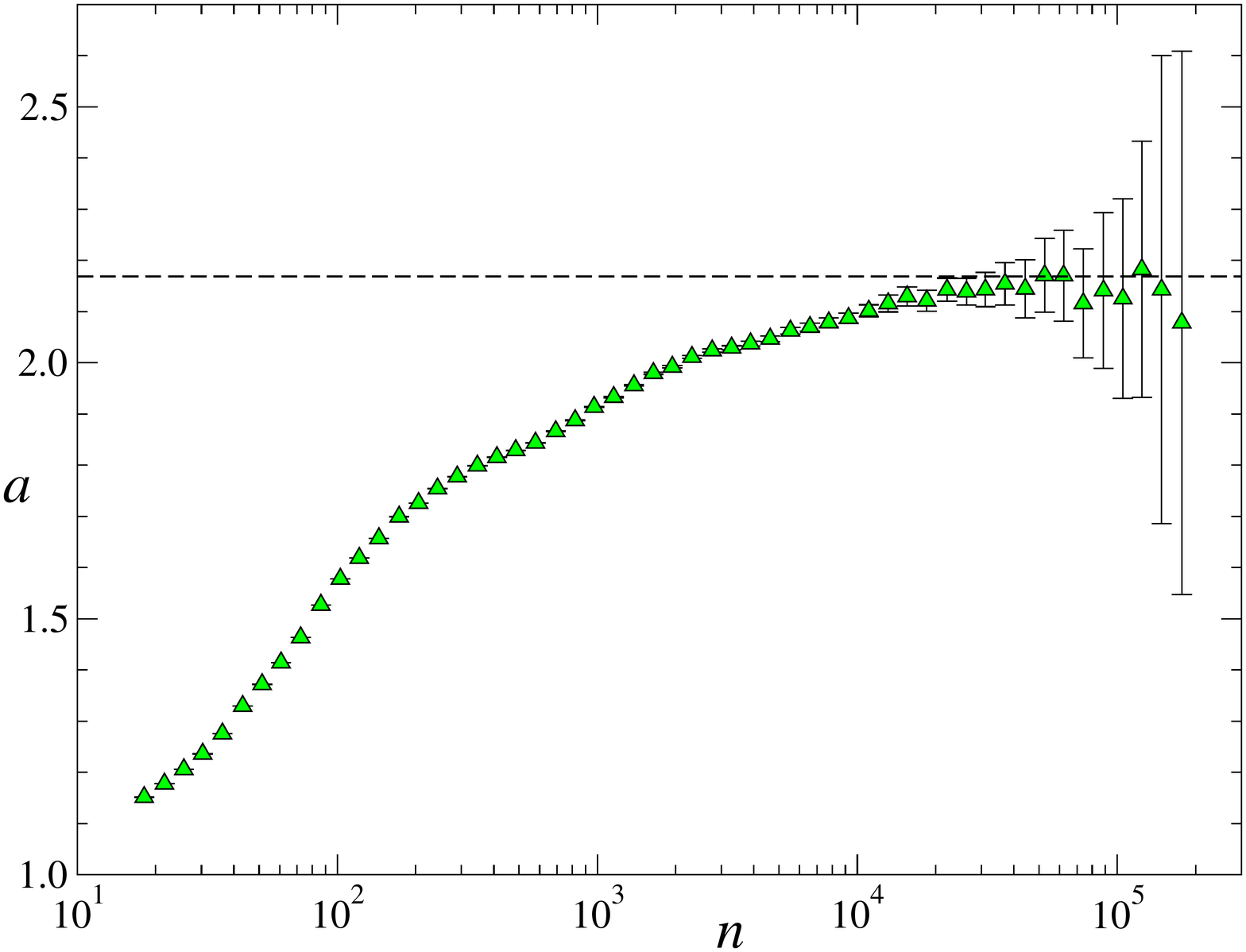}
  \caption{Results of a least-squares fit of $p_n=a n^{-\delta}$ to a
    sliding window of $w=10$ consecutive data points. The abscissa represents
    the geometric mean of the $n$-values contained in a window.}
  \label{fig:fit-parameters}
\end{figure}

The numerical value of the fit parameter varies with the choice of the
data points used in the least square fit, however. As a more
systematic way to estimate the asymptotic behavior we applied least
squares fitting of the two-parameter function $a n^{-\delta}$ to
sliding windows of $w$ consecutive data points
$p_n$. Figure~\ref{fig:fit-parameters} shows the results for $w=10$.
Larger values of $w$ yield similar curves with smaller errorbars but
fewer data points. This analysis shows that the available numerical
data for $p_n$ supports the conjecture \eqref{eq:conjecture} within
the errorbars.

A more elementary question is whether $\lim_{n\to\infty} p_n$ is zero
or non-zero.  To address this question we applied the sliding window
technique to fit the three parameter function $p_n=a n^{-\delta} +
b$. The result for $b$ is a curve very similar to the curves shown in
Figure~\ref{fig:fit-parameters} with $b$ converging to zero within the
errorbars. This result supports the claim $\lim_{n\to\infty} p_n = 0$.

\section{Conclusions and Outlook}

We have demonstrated that Irving's algorithm for the stable roommates
problem can be organized such that the expected time and space
complexity is $O(n^{3/2})$ on random instances. Our reasoning about
the dynamics of the algorithm (approaching random walks of leftrank
and rightrank, phase I as coupon collector's problem) is of course
non-rigorous, but the results are well confirmed by the numerical
simulations. Maybe this simplistic view on Irving's algorithm can help
to derive the observed $n^{-1/4}$ decay of the probability $p_n$ that a
random instance of size $n$ has a solution.

\ack{%
  I am grateful for comments from David Manlove and Ciaran McCreesh.
}

\section*{References}

\bibliographystyle{unsrt} 
\bibliography{mertens,matchings}

\end{document}